\documentclass[12pt]{iopart}
\usepackage{graphicx}

\begin{document}
\title[Lithium-Beryllium Method]{A Lithium-Beryllium Method for the Detection of Solar
Neutrinos}

\author{A.V.Kopylov, I.V.Orekhov, V.V.Petukhov, and A.E.Solomatin}

\address{Institute of Nuclear Research of Russian Academy of Sciences \\
117312 Moscow, Prospect of 60th Anniversary of October Revolution
7A, Russia}

\ead{beril@al20.inr.troitsk.ru}

\begin{abstract}A method for the detection of solar neutrino has been developed
using the laboratory bench installations. The efficiency of the
extraction of beryllium from lithium as high as 96.4{\%} has been
achieved, and it was shown that lithium losses during the extraction
were less than 1{\%}. The prospects of a full-scale experiment with
a 10-t lithium detector consisting of twenty 500-kg lithium modules
are discussed. The technical solutions formulated on the basis of
this study enable to make design of a pilot lithium installation
containing 500 kg of metallic lithium.
\end{abstract}
\noindent{\it neutrino experiments}

\maketitle

\section{Introduction}
The use of $^{7}$Li isotope as a promising target for capturing
solar neutrinos was first proposed by Zatsepin and Kuzmin \cite{1}
and independently by Bahcall \cite{2}. Nuclei $^{7}$Li and $^{7}$Be
are the mirror nuclei, and transitions between them are
superallowed. Therefore, the cross section of the reaction

\begin{equation}
^7Li + \nu _e  \to  ^7Be + e^{-}
\end{equation}

is relatively large and can be calculated with high accuracy. Due to
the thermal broadening of the energy spectrum of $^{7}$Be neutrinos,
the detector has a certain sensitivity to beryllium neutrinos, as
was first shown by Domogatsky \cite{3} and later was confirmed more
accurately by Bahcall \cite{4}. Other factors that make the detector
very attractive are the small atomic weight of the target and a high
(92.5{\%}) abundance of $^{7}$Li in natural lithium. Taken together,
these factors provide a small mass of a lithium target with a
relatively high sensitivity to solar neutrinos of medium energy
range. Ten tons of lithium suffice to construct a full-scale solar
neutrino detector. Some specific points concerning the
implementation of the lithium project were considered elsewhere
\cite{5} - \cite{11}.

\section{The General Analysis Of The Method}
At present the statement can be made that a radiochemical
lithium-beryllium method with a metallic lithium target to detect
solar neutrinos has been developed in principal questions of its
implementation for a full-scale experiment. The main advantage of
the lithium detector is a high sensitivity to neutrinos generated in
the carbon-nitrogen-oxygen (CNO) cycle. The radiochemical technique
was chosen because electronic detectors meet serious problems with
the detection of these neutrinos: the fluxes are weak and energies
are relatively low. The advantage of metallic lithium as a target is
that some background-producing channels (fragmentation of the target
by the nuclear-active component of cosmic rays and background from
the natural radioactivity) are either absent or much weaker than
those in case of lithium compounds \cite{12}. To implement the
lithium project, it is of prime importance to develop (i) a
technique for the extraction of beryllium from metallic lithium and
(ii) a method to count the $^{7}$Be atoms extracted from lithium.
The $^{7}$Be isotope is rather difficult to detect, because the
kinetic energy of Auger electrons released as a result of its decay
is only 55 eV. This is -- the main problem for the construction of a
low-background counting system. However, this problem can be
overcome if to use $^{7}$Be decays to the excited level of $^{7}$Li.
The BR (Branching Ratio) of this channel of decay is low (only 10.3
{\%}). However, it was shown by calculations \cite{13} that it
suffices to achieve an accuracy of the measurement of the effect
from solar neutrinos with the uncertainty approximately 10{\%} what
is, in fact, the aim of a lithium experiment. Since a gamma quantum
with the energy of 478 keV is emitted in this case, i.e. one has a
rather convenient energy for counting by means of HPGe detector, the
design of an appropriate counting system on the basis of advanced
technologies is quite feasible. Using an assembly of high-purity
germanium detectors with the 4$\pi$ geometry, the efficiency of
$^{7}$Be detection of approximately 8{\%} can be achieved with a
sufficiently good background discrimination. This method of
detection is also convenient because one can use a solid sample
containing extracted $^{7}$Be atoms up to a few grams. It is
important to prevent degradation of the counting efficiency due to
the self-absorption of the 478-keV gamma line, this is the main
limitation for the weight of the sample. The use of a solid sample
considerably eases the requirements regarding the technology of
lithium extraction from beryllium as will be explained later on.

Metallic lithium is used in a number of applications as a good
heat-carrying agent. This point has been touched upon in many
publications (and in scientific and technical conferences). Here, we
concentrate on the following issues. The technology of closed
liquid-metal lithium loops, which is employed today, is of little
use for the lithium project, since the design of these loops is
rather sophisticated. The lithium detector must work for a
sufficiently long time (5 years). Moreover, emergency cases should
be practically excluded by two reasons: first -- because lithium
detector should be placed in a deep underground laboratory and
second -- to minimize the dead time of the detector. The design of
the facility must be as simple and reliable as possible. The
efficiency of beryllium extraction from lithium must be no less than
90{\%}; otherwise, the overall efficiency of the system will
substantially decrease. And the most important point - after each
procedure of beryllium extraction, the system must recover to its
initial state; i.e., the characteristics of the detector must not be
changed in time. It is also necessary to minimize lithium losses as
a result of the extraction procedure. Such are the main requirements
to the detector under development.

\section{Experimental}
The technology was tried out on different laboratory benches. At the
first stage, on the way to simplify the technology, the important
result was obtained: beryllium can be extracted from lithium at
thermostatic conditions; i.e., there is no need to vary the
temperature during the extraction procedure. This substantially
simplifies the design of the facility and decreases the energy
consumption. The next problem was to validate the chosen technology.
For this a special setup was assembled shown in Figure 1.

\begin{figure}[!ht]
\centering
\includegraphics[width=3in]{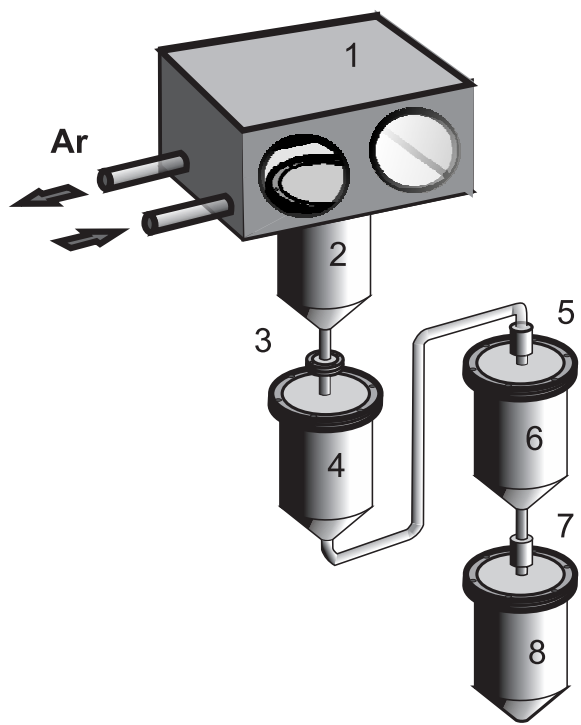}
\caption{(1) Argon-filled box, (2) feed cylinder no.1, (3) device to
download aluminum, (4) drain cylinder no. 2, (5) filter no.1, (6)
cylinder no.3, (7) filter no.2, and (8) cylinder no.4}
\end{figure}

A stainless steel cylinder (2) is placed in a glove box (1) through
which argon has been swept. High-purity (99.9{\%}) lithium of 3.2 l
(1.6 kg in mass) was placed into this cylinder in an argon
atmosphere through the upper lid with a flanged joint. A stainless
steel mesh placed at the bottom of the tank was used to remove films
of the lithium oxide. Through a pipe at the bottom of the cylinder
the molten lithium was delivered to assembly (3), which contained 50
mg of high-purity aluminum irradiated by a $\sim$ 100-MeV proton
beam on the high-intensity proton accelerator (Institute for Nuclear
Research of the Russian Academy of Sciences, Troitsk, Moscow
Region). The irradiation produced a certain activity of $^{7}$Be and
$^{22}$Na in the aluminum sample. This pair of isotopes turned out
to be very convenient to study the extraction of beryllium from
lithium.

Figure 2 shows the energy spectrum of the gamma radiation from the
irradiated sample that was obtained using 100 cc HPGe detector.

\begin{figure}[!ht]
\centering
\includegraphics[width=4in]{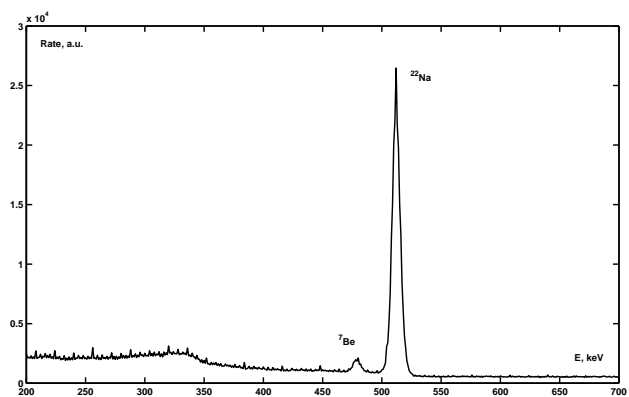}
\caption{Energy spectrum of gamma radiation from the irradiated
aluminium sample}
\end{figure}

One can see that activity of $^{22}$Na was much higher than that of
$^{7}$Be. This fact might cause some problems in the subsequent
measurements because the low energy tail of the sodium line 511 keV
might mask the beryllium line 478 keV. Therefore, the aluminum
sample was purified from sodium by evaporating $^{22}$Na atoms
heating the aluminum sample in a graphite crucible in a vacuum at
$800^{\circ}C$ till the intensities of these isotopes became
comparable.

Figure 3 shows the gamma spectrum of the sample measured after the sodium
evaporation.

\begin{figure}[!ht]
\centering
\includegraphics[width=4in]{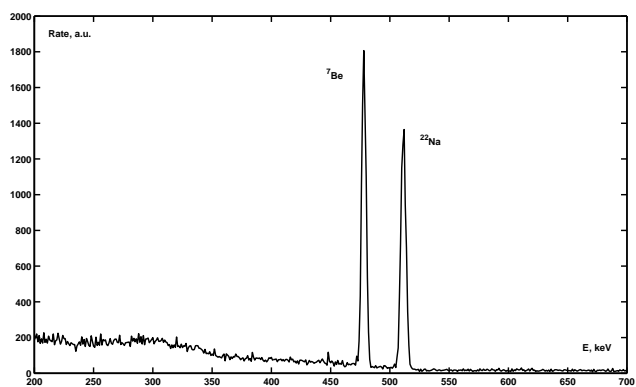}
\caption{Energy spectrum of gamma radiation from the irradiated
aluminium sample after elimination of $^{22}$Na}
\end{figure}

It was very interesting to see whether or not beryllium deposits
onto the aluminum oxide film during melting. If so, the distribution
of the beryllium over the lithium volume could be not homogeneous
after the aluminum is introduced, as a carrier, into the lithium
melt. To check this, the aluminum oxide film was etched off in
hydrofluoric acid. Measurements of the sample activity after the
etching showed that the beryllium was not deposited on the oxide
film. This means that this way of the beryllium insertion does
provide homogeneous distribution of beryllium in lithium. The
aluminum and sodium were also uniformly distributed over the lithium
volume, since their solubility in lithium is high. Then lithium
containing $^{7}$Be and $^{22}$Na activity was delivered into a
cylinder no. 3 through filter no. 1 with three tight-wound meshes
and then into a cylinder no. 4 through filter no. 2. During the
extraction procedure, all the components of the setup were kept at
$210^{\circ}C$. Then the system was cooled down to room temperature
and was cut into parts and the activity of each part was measured
using a gamma spectrometer. The measurement of the activity of each
part has revealed the following:

\begin{enumerate}
\item The $^{22}$Na isotope almost entirely entered into the lithium contained in a cylinder no. 4. The losses of this isotope (deposits on the walls and filters) were less than 1{\%}.
\item No traces of $^{7}$Be were observed in the lithium contained in a cylinder no. 4; that is, beryllium has been completely extracted from lithium.
\item No traces of $^{7}$Be were observed on the walls of cylinder no. 3 and on filters nos. 1 and 2. Less than 1{\%} of the beryllium escaped the cylinder no.2 (drain cylinder).
\item Almost all the activity of $^{7}$Be was deposited on the oxide films that were left on the walls of the drain cylinder. Partly this has been achieved due to a very small diameter of the discharge pipe at the outlet of this cylinder. (Would be the diameter of this tube larger, some portion of the oxide film with the corresponding activity of $^{7}$Be would inevitably flow away to a filter no.1).
\end{enumerate}

The deposition of $^{7}$Be on the walls facilitates the subsequent
extraction procedure (it is much easier to collect $^{7}$Be from the surface
than from the volume of the filter) and also minimizes lithium losses in the
extraction procedure. The atoms of $^{7}$Be contained in the oxide films on
the walls of a drain cylinder were flushed away by subsequent showering the
walls by measured portions of water and alcohol mixture. After flushing the
walls of cylinder, each portion of water was passed through a paper filter.
As a result, part of the activity of $^{7}$Be was in an insoluble residue on
the filter and part - in the solution downstream of the filter.

A total of five flushing, 100 ml of water in each, were made.
Activity measurements taken after each flushing showed that the
first four flushes wash away 96.4{\%} of the initial activity of
$^{7}$Be, 61.4{\%} of which was in the precipitate and the remaining
35{\%} in the soluble phase. An amount of 96.4{\%} is quite an
acceptable result for a lithium experiment. Next, after burning
filters and measuring the activity before and after burning, we
found that the activity of $^{7}$Be was not lost in this procedure.
This is because beryllium has no volatile compounds. Thus, the
efficiency of such a technique for beryllium concentration was
demonstrated. The beryllium was extracted from the solution and
converted to the solid phase using a technique developed earlier
\cite{14}. Losses of the activity of $^{7}$Be on the level more than
1{\%} were not observed. The measurement of the lithium content in
the solution showed that lithium losses in the process of beryllium
extraction were less than 1{\%}, what is quite satisfactory to
conduct a full scale experiment.

\section{Results and Discussion}
The results of this work can be summarized as follows.

\begin{enumerate}
\item Extraction of beryllium from lithium must proceed at thermostatic conditions.
\item A modular structure (20 modules 500-kg lithium each) of the detector is optimal.
\item Each module consists of two thermally insulated tanks shearing a common case, as shown in Fig. 4. The lithium is transferred back and forth from one tank of the module to the other. To facilitate the collection of beryllium atoms on the film of oxides on top of the molten volume, lithium gets mixed for a few hours at the end of exposure.
\item After the lithium is transferred to another tank, the beryllium is collected from the precipitate on the walls of the tank by flushing the walls.
\item The alkaline lithium solution is filtered to obtain an insoluble residue and a solution.
\item Beryllium is extracted from aqueous solution, purified and concentrated until a solid residue is obtained which can be used as a sample for the assembly of HPGe detectors.
\end{enumerate}

\begin{figure}[!ht]
\centering
\includegraphics[width=3in]{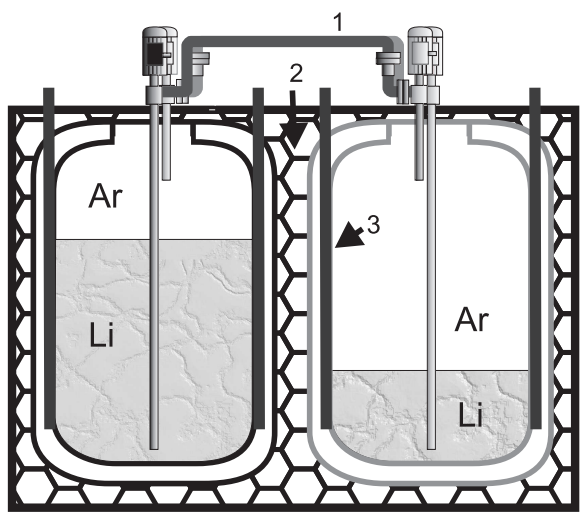}
\caption{Schematic of the 500 kg (1 m$^3$) lithium module: (1)
liquid metal tubes and valves, (2) thermal insulation, and (3)
electric heaters}
\end{figure}

These are the major steps of the technological process which can be
put at the basis for design and construction of a pilot lithium
module.

One of the background sources of lithium detector is the intrinsic
radioactivity from uranium and thorium isotopes in lithium. The
concentrations of uranium and thorium (which were taken to be in
equilibrium with daughters) in lithium were measured using a
low-background NaI(Tl) detector surrounded by a 30-kg lithium layer.
The measurements lasted 40 days with the lithium layer around the
detector; then the background was measured for 60 days when lithium
sample was removed. Upon the results of data treatment the upper
limits were found of the abundance of $^{238}$U - 4 x 10$^{-10}$
g/g, and of $^{232}$Th - 2.2 x 10$^{-10}$ g/g. After the systematic
uncertainties taken into account, an upper limit 10$^{-9}$ g/g both
for $^{238}$U and $^{232}$Th can be taken with high (90{\%})
confidence. At this level the calculated $^{7}$Be production rate
from the impurities of $^{238}$U and $^{232}$Th is less than 1{\%}
of the effect expected from solar neutrinos, which is quite
satisfactory. The background was estimated using computational
results reported in \cite{12}.

\section{Conclusion}
The optimal conditions for extraction of beryllium atoms from metallic
lithium are found, the coefficient of extraction 96{\%} is obtained. The
loss of lithium during the extraction procedure has shown to be less than
1{\%}. Prior to the construction of a full-scale 10-t lithium installation,
it is planned to check all technological aspects on a pilot one-module setup
where the reliability of the method can be demonstrated.

Basically, the lithium experiment can provide important information
on the thermonuclear reactions in the Sun and together with the
results of other solar neutrino experiments be used to further study
the properties of neutrinos. At present, the development of the
technique of lithium-beryllium experiment is quite timely, because
the sensitivity of electronic detectors under development is still
insufficient for reliable detection of neutrinos produced in the CNO
cycle. The results of this study have shown that principally it is
possible to realize a full-scale experiment with 10 t of metallic
lithium using a setup consisting of 20 lithium modules. This setup
can be placed at a connection of two tunnels at a depth of about
4000 hg/cm$^{2}$ and does not require excavation of a separate
chamber, which simplifies the project and makes it much cheaper.
Further efforts must be focused on designing a 500-kg one-module
pilot facility and a counting system based on an assembly of
high-purity germanium detectors.

The R{\&}D work performed here on metallic lithium has also a
certain value for the applied physics. The technology, conditions,
and geometry for efficient extraction of beryllium from lithium have
been found. This can be used, in particular, for creating intense
neutron sources based on a lithium target (the IFMIF project
\cite{15} is an example).

\section{Acknowledgments}
The authors are grateful to G.T.Zatsepin, V.A.Kuzmin, and
L.B.Bezrukov for fruitful discussions, to G.Ya.Novikova for the
assistance in measuring lithium content in an aqueous solution and
to B.L.Zhuikov for the assistance in the irradiation of aluminum
sample by a high-intensity proton beam. This work was supported by
the Russian Federation Program in Support of Leading Scientific
Schools (grant no. LSS-1786.2003.2), the basic research program
Neutrino Physics at the Russian Academy of Sciences, and the Russian
Foundation for Basic Research (grant no. 04-02-16678).


\section*{References}
\begin{thebibliography}{99}

\bibitem{1} V.A.Kuzmin and G. T. Zatsepin, in {\it Proceedings of the 9th International Cosmic Ray Conference} (1965) 1024
\bibitem{2} J.N.Bahcall, {\it Phys.Rev.Lett.} {\bf 13} (1964) 332; {\it Phys.Rev.Lett.} {\bf 23}
(1969) 251
\bibitem{3} G.V.Domogatsky, Preprint No. 153, FIAN (Lebedev Physical Institute, Moscow) (1969)
\bibitem{4} J.N.Bahcall, {\it Rev.Mod.Phys.} {\bf 50} (1978) 881
\bibitem{5} J.K.Rowly, in {\it Proceedings of the Conference on Status and Future of Solar Neutrino Research} (BNL) USA (1978) 265
\bibitem{6} E.Veretenkin, V.Gavrin, and E.Yanovich, {\it Russ.J.At.Energy} {\bf 88}
(1985) 65
\bibitem{7} A.V.Kopylov, A.N.Likhovid, E.A.Yanovich, and G.T.Zatsepin, in {\it Proceedings of the International School "Particles and Cosmology"} Baksan Valley  Russia,
(1993) // Kh.S.Nirov, and V. A. Rubakov {\it World Sci.}, Singapore,
1993)63
\bibitem{8} S.Danshin, G.Zatsepin, A.Kopylov, et al., {\it Phys.Elem.Part. and Atom.Nucl.} {\bf 28} (1997) 5
\bibitem{9} M.Galeazzi, G.Gallinaro, F.Gatti, et al., {\it Phys. Lett. B} {\bf 398} (1997) 187
\bibitem{10} M.Galeazzi, G.Gallinaro, F.Gatti, P.Meunier, S.Vitale,
and A.Kopylov, and E.Yanovich, {\it Nucl. Instrum. Methods Phys.
Res. A} {\bf 401} (1887) 317
\bibitem{11} A.V.Kopylov, I.V.Orekhov, V.V.Petukhov, and A. E. Solomatin, {\it Phys.Elem.Part. and Atom.Nucl., Lett.} {\bf 4} (2007) 552
\bibitem{12} Yu.I.Zakharov, {\it Proceedings of the International Conference "Neutrino Physics and Neutrino
Astrophysics" (Neutrino-77)} Baksan Valley, Russia (1977) {\bf 1} 64
\bibitem{13}  A.Kopylov and V.Petukhov, {\it arXiv:hep-ph/0301016}, {\it
arXiv:hep-ph/0306148}
\bibitem{14}  G.Ya.Novikova, {\it PhD thesis} (2001).
\bibitem{15}  {\it IFMIF-KEP: International Fusion Materials Irradiation Facility-Key Element Technology Phase}
Rep.Jpn. Atomic Energy Res. Inst. (2003).
\end{thebibliography}
\end{document}